\documentclass{mn2e}
\usepackage{epsfig}

\def\le{{L_{\rm Edd}}}
\def\msun{{\rm M_{\odot}}}

\def\me{{\dot M_{\rm Edd}}}
\def\va{{v_{\rm A}}}
\def\vk{{v_{\rm K}}}
\def\pb{{p_{\rm B}}}
\def\cs{{c_s}}
\def\cg{{c_g}}

\title[Strong Fields and Discs]
{Accretion Discs with Strong Toroidal Magnetic Fields}

\author[M.C. Begelman \& J.E. Pringle]
{M. C. Begelman$^{1,2}$ \& J.E. Pringle$^{1,2,3}$ 
\\
$^1$JILA, University of Colorado, Boulder, CO 80309-0440, USA\\
$^2$Institute of Astronomy, Madingley Road, Cambridge, CB3 0HA\\
$^3$JILA Visiting Fellow\\
}

\begin{document}

\maketitle
\begin{abstract}
Simulations and analytic arguments suggest that the turbulence driven
by magnetorotational instability (MRI) in accretion discs can amplify
the toroidal (azimuthal) component of the magnetic field to a point at
which magnetic pressure exceeds the combined gas + radiation pressure
in the disc.  Arguing from the recent analysis by Pessah and Psaltis,
and other MRI results in the literature, we conjecture that the
limiting field strength for a thin disc is such that the Alfv\'en
speed roughly equals the geometric mean of the Keplerian speed and the
gas sound speed.  We examine the properties of such
magnetically-dominated discs, and show that they resolve a number of
outstanding problems in accretion disc theory.  The discs would be
thicker than standard (Shakura-Sunyaev) discs at the same radius and
accretion rate, and would tend to have higher colour temperatures.  If
they transport angular momentum according to an $\alpha$ prescription,
they would be stable against the thermal and viscous instabilities
that are found in standard disc models.  In discs fuelling active
galactic nuclei, magnetic pressure support could also alleviate the
restriction on accretion rate imposed by disc self-gravity.
\end{abstract}


\section{Introduction}

The standard theory of geometrically thin accretion discs (Shakura \&
Sunyaev 1973) has provided a successful framework for understanding
the basic physical properties of these objects. And good progress
toward understanding the main uncertainty in the theory, the
viscosity, has been made since it was realized that the
magneto-rotational instability (MRI) can drive dynamo action to
maintain magnetic stresses (Balbus \& Hawley 1998).  However, several
aspects of disc theory have remained problematic over the years.

Four of these problems stand out. First, observed disc spectra in cataclysmic variables (CVs) deviate significantly from theoretical expectations, even when self-irradiation is taken into account.  Moreover, different kinds of CV systems with similar accretion rates often show different spectra. Second, the discs in CVs and low-mass X-ray binaries (LMXBs)
often appear to be geometrically thicker than expected, suggesting
that some additional source of pressure supports the optically thick
layers against gravity, over and above the usual mix of gas +
radiation pressure. Third, disc models in which the viscous stress is
roughly proportional to the total pressure (``$\alpha$-models") --- an
otherwise plausible assumption --- are often predicted to be unstable
to perturbations in temperature and surface density.  Yet observations
of many potentially unstable sources do not show clear evidence of
such instabilities. Fourth, the accretion discs thought to fuel active
galactic nuclei (AGN) are predicted to be gravitationally unstable on
scales exceeding a fraction of a parsec, yet mass transport does not
seem to be disrupted by fragmentation and star formation in many AGN.
 
In this paper, we propose a mode of disc accretion in which the
pressure of the organized toroidal magnetic field dominates over other
forms of pressure.  On the basis of published numerical simulations,
we argue that the MRI-driven dynamo could amplify the toroidal field
to a strength that greatly exceeds the gas pressure.  Similar
arguments were presented by Pariev, Blackman \& Boldyrev (2003) and
Blaes et al.~(2006), but the former focused on the stochastic part of
the field, and neither attempted to estimate the field strength from
first principles.  Arguing on the basis of analyses by Pessah \&
Psaltis (2005) and others, we suggest that MRI-driven turbulence could
amplify the toroidal field to a maximum strength such that the
associated Alfv\'en speed is roughly the geometric mean between the
Keplerian speed and the gas sound speed.  Such a strong field can
dominate the disc structure even in regions where radiation pressure
dominates over gas pressure.  We construct a model of a magnetically
dominated $\alpha$-disc according to this prescription, and show that
the additional pressure support both thickens the disc and increases
its colour temperature, stabilizes thermal and viscous instabilities,
and ameliorates the effects of gravitational instability.

The plan of the paper is as follows. In Section 2, we outline the
evidence for anomalous spectra and disc thicknesses in CVs
and LMXBs.  We defer brief discussions of the stability problems to
Section 4.  In Section 3, we discuss the numerical
evidence supporting strong toroidal fields in discs, then focus on the
Pessah \& Psaltis (2005) analysis and results, as well as results in
the literature for nonaxisymmetric MRI, in order to justify our
estimate for the characteristic disc field. In Section 4 we adopt
eq.~(\ref{pb}) as the constitutive relation for the magnetic
pressure, and construct a family of simple black hole disc models
characterized by an $\alpha-$viscosity under this assumption. We show
that the resulting discs are hotter and thicker than standard
discs. Most importantly, and in contrast to the inner regions of
standard black hole disc models, they are not subject to thermal or
viscous instabilities. In addition, in the outer regions of AGN discs,
where self-gravity is important, they can carry much larger mass
fluxes without fragmenting due to gravitational instability.  Finally,
we discuss our results and summarize our conclusions in Section 5.

\section{Problems with CV and LMXB disc models}

\subsection{Spectral Properties}

In principle a steady-state, geometrically thin accretion disc should be relatively
simple to understand. Since the accretion rate is independent of
radius, $R$, the effective temperature $T_{\rm eff}(R)$ is independent
of the viscosity mechanism, and is given just in terms of system
parameters (Pringle 1981) as
\begin{equation}
T_{\rm eff}(R) = T_* \left( \frac{R}{R_*} \right)^{-3/4} \left[ 1 - \left( \frac{R}{R_*} \right)^{-1/2} \right]^{1/4},
\end{equation}
where $R_*$ is the radius of the inner disc edge, and the reference
temperature $T_*$ is given in terms of the mass of the central object,
$M_*$, and the accretion rate, $\dot{M}$, by
\begin{equation}
T_* = \left( \frac{ 3 G M_* \dot{M} }{8 \pi \sigma_{\rm SB} R_*^3 }
\right)^{1/4},
\end{equation}
where $\sigma_{\rm SB}$ is the Stefan--Boltzmann constant.

The astronomical objects to which this formula should be most
straightforwardly applicable are the accretion discs in cataclysmic
variables (CVs), and in particular the discs in nova-like
variables, and dwarf novae in outburst. In the nova-like variables,
and in the Z Cam variables in standstill, the discs are steady-state
in that the timescale for variation of the accretion rate (measured by
system brightness) is much longer than the viscous timescale (as
deduced empirically from the evolution timescale for dwarf nova
outbursts). Indeed, even in dwarf novae in outburst, and especially in
super-outburst, the discs are close enough to being steady state that
their spectra should be readily calculable. Estimates of accretion
rates and evolution timescales imply that these discs are optically
thick. Thus, if the accretion energy is deposited within the body of
the disc (where most of the matter resides), computing the spectra
should be closely akin to the well understood problem of computing the
spectra of stellar atmospheres, especially given that in this case the
temperatures and gravities (densities) involved are in the standard
range (Wade \& Hubeny 1998).

However, it has long been known that this simple-minded approach does
not give satisfactory results. Just from an empirical point of view it
was already clear that there might be problem: in her review of IUE
spectra, la Dous (1991) showed that in the range $\lambda \lambda 1200
- 3000$ the continuum spectra of dwarf novae at the peak of an
outburst are significantly ``bluer" (i.e. have more flux at shorter
wavelengths) than the continuum spectra of non-magnetic nova-like
stars. For the nova-like variables, which should most closely
approximate steady-state discs, it turns out that it is difficult, if
not impossible, to replicate the UV data using steady-state disc
models constructed from LTE stellar atmospheres (Wade 1988; Long et
al.~1994; Orosz \& Wade 2003). In general, the models predict UV
spectra that are too ``blue," and in addition it is not possible to
match both the observed flux and colour at the same time. Similar
problems are found for dwarf novae at the peak of outburst (Knigge et
al.~1997)

\subsection{Disc thickness}

In the standard accretion disc picture, the thickness of the disc is
governed by a balance between pressure and (a component of)
gravity. To a first approximation, the scaleheight of the disc, H, is
given by
\begin{equation}
\label{honr}
H \sim R \left( \frac{c_s}{V_\phi} \right)
\end{equation}
(Pringle 1981), where $c_s$ is the sound speed at the disc midplane
and $V_\phi$ is the azimuthal velocity. For a simple steady-state CV
disc, the shape of the disc is slightly flared in that $H/R \propto
R^{1/8}$ (Shakura \& Sunyaev 1973).

In order to fit the eclipse profiles of discs in CVs, it is necessary
to allow for the fact that the disc is not flat, and for the fact that
the disc photosphere is usually at a height of a couple of disc
scaleheights above the disc midplane. Eclipse fitting does allow some
estimation of the disc thickness, usually in terms of the value of
$H/R$ at the outer disc edge. The observationally inferred values of disc
thickness required to fit eclipse observations are typically a couple of
times larger than the theoretical estimates (Robinson, Wood \& Wade
1999; Shafter \& Misselt 2006).

There are other indications of material farther out of
the plane than simple estimates would suggest, notably the ``iron
curtain" material seen both in quiescent discs (Horne et al.~1994) and
at high mass transfer rates (Baptista et al.~1998), and transient
non-axisymmetric features, possibly spiral structure seen in tenuous
layers of the atmosphere (Steeghs, Harlaftis \& Horne 1997), which may
require an unusually hot, or thick, component of the atmosphere and/or
additional tidal heating at the disc edge (Ogilvie 2002).

LMXBs (White \& Mason 1985; Hakala, Muhli \& Dubus 1999) and the Algol
binary W Cru (Pavlovski, Burki \& Mimica 2006) provide more evidence
that the outer disc edge does not fit simply with standard disc
theory, in that the outer disc edge seems to be thick and structured.
In addition, the apparent paucity of eclipses among LMXBs, or bulge
sources, indicates that the disc manages to obscure the view of the
central X-ray source from a larger fraction of the sky than might be
expected (Milgrom 1978; Joss \& Rappaport 1979).

The disc rim may not be the ideal place to test standard disc theory,
since the outer disc edge is subject to both tidal dissipation and the
impact of the accretion stream (Lubow 1989; Armitage \& Livio 1998).
Nevertheless, the soft X-ray transients do seem to indicate that in
these objects the disc thickness is larger than expected at most
radii, and not just at the outer disc edge.  This is because, in order
to explain the prolonged nature of the outbursts and the shape of the
decay lightcurve, the simplest explanation appears to be that the disc
is significantly irradiated from the centre (King 1998). According to standard disc models, such
irradiation is not possible, since the inner parts of the disc screen the outer regions (Dubus et
al.~1999). However, one can provide an adequate description of the outbursts of soft X-ray transients by assuming that the
disc is thicker than simple models predict, allowing significant irradiation to occur (Dubus, Hameury \& Lasota 2001).

\section{How Strong are Disc Magnetic Fields?}

It seems likely that the problems discussed above (and several to be
discussed later) could be resolved by considering the dynamical
effects of magnetic fields in accretion discs. In standard disc
theory, spectral predictions are based on two assumptions: first, that
energy is deposited where most of the mass is, and second, that the
density structure of the disc is smooth and homogeneous.  However, it
has long been realized that if the process that taps the shear energy,
first converts it to magnetic energy, which is then dissipated, then
much of the energy can be deposited in a small fraction of the mass
(Lynden-Bell 1969; M\'esz\'aros, Meyer \& Pringle 1977).  Moreover, it
has also long been known that if a disc is strongly magnetic, it is
likely to be highly inhomogeneous (Pustilnik \& Shvartsman 1974). In
recent models of spectrum formation in the low/hard state of discs in
LMXBs and for the hard X-ray flux in AGN, these ideas are taken to the
limit in which all the accretion energy is assumed to be deposited in
a low density, extended magnetic corona (see, for example, $\dot{\rm
Z}$ycki, Done \& Smith, 2001; Done \& Nayakshin, 2001, Barrio, Done \&
Nayakshin, 2003; and references therein).

Numerical simulations give some support to these ideas. Hirose, Krolik
\& Stone (2006) and Fromang \& Nelson (2006), following on from
earlier work by Miller \& Stone (2000), present the results of
shearing-box simulations of MRI-driven accretion discs, in which the
vertical extent of the computational grid is large enough to encompass
regions of low gas density. The typical structure that they find is a
magnetic sandwich, with most of the mass forming a gas
pressure-dominated disc in the plane, but with extensive, low density,
magnetically supported layers on either side. In these simulations,
most of the energy dissipation still occurs in the gas
pressure-dominated central layer, and therefore at high optical
depth. However, the optical depth of the extended magnetic atmosphere
is non-negligible, and this has two effects. First, the density at the
effective photosphere is lowered, increasing the ratio of scattering
to absorptive opacity, and so giving rise to a slightly harder
spectrum (Blaes et al. 2006). Second, the height of the apparent
photosphere is increased, making the disc appear thicker than a
standard disc.
 
In this paper we raise the possibility of there being a second
equilibrium distribution for the magnetic flux in a disc, namely one
in which the magnetic pressure ($p_B = B^2/8\pi$) dominates at all
heights in the disc (or at least through most of the pressure scale
height).  The existence of such a distinct class of discs has been
raised before by Shibata, Tajima \& Matsumoto (1990). As we discuss
below, such a possibility would be difficult to simulate numerically
because of the small lengthscales of the instabilities that maintain
the magnetic dynamo.  It is also likely that the formation of such a
strongly magnetized disc requires special conditions of some kind. For
example, Pringle (1989) discusses the formation of a strongly
magnetized region in an accretion disc boundary layer, where the
strong shear leads to a much higher formation rate for toroidal
flux. A disc might also become strongly magnetically dominated as a
result of thermal instability, for example in the central regions of
black hole accretion discs (Machida, Nakamura \& Matsumoto 2006), or
during the transit to quiescence in a cataclysmic variable disc (Tout
\& Pringle 1992). In these circumstances the disc is originally hot,
with $p_B \sim \alpha p_g$, where $p_g$ is the gas pressure and
$\alpha$ is the Shakura-Sunyaev (1973) viscosity parameter. The disc
then cools (reducing $p$ but not $p_B$) rapidly on a thermal timescale
($\sim 1/\alpha \Omega$, where $\Omega$ is the angular speed). This
would result in a disc in which $\beta = p_g/p_B \ll 1$.

Pariev et al.~(2003) have considered the structure of a strongly
magnetic disc in which $\beta \ll \alpha \sim 1$, but were unable to
predict the strength of the field. In this paper, we attempt to
quantify the characteristic strength of the toroidal magnetic fields
likely to dominate accretion discs, and to explore the consequences of
these strong fields for vertical and radial disc structure.  We base
our estimate of the limiting field strength on a recent analysis of
the effects of strong toroidal fields on MRI by Pessah \& Psaltis
(2005), as well as other analyses in the literature.  We assume that
toroidal fields are amplified by dynamo action resulting from the
turbulence driven by MRI, and are probably limited by buoyancy effects
operating on some multiple of the dynamical (orbital) timescale.  As
long as MRI grows on a dynamical timescale, the azimuthal field,
$B_\phi$, continues to be amplified, but when the MRI growth rate is
suppressed by magnetic tension the growth stops and $B_\phi$
saturates.

MRI continues to operate on a dynamical timescale even when the
toroidal magnetic pressure exceeds the gas pressure. Pessah \& Psaltis
(2005) show that the maximum growth rate for axisymmetric MRI modes
begins to be severely suppressed only when the Alfv\'en speed
associated with the toroidal field, $v_{{\rm A}\phi} = (B_\phi^2 /
4\pi\rho)^{1/2}$, exceeds the geometric mean of the Keplerian speed
$v_{\rm K}$ and the gas sound speed $c_g$. (Radiation pressure is not
considered in their analysis; we address this later.)  When $v_{{\rm
A} \phi} = (2 c_g v_{\rm K})^{1/2}$, the growth rate of MRI vanishes.
Although different types of axisymmetric and nonaxisymmetric modes may
grow when the field strength is larger than this limit, we conjecture
the $v_{{\rm A} \phi}$ cannot grow past this instability
``bottleneck."  We therefore adopt this limiting value for
axisymmetric MRI as a measure of the characteristic magnetic pressure,
at least when gas pressure dominates:
\begin{equation}
\label{pb}
p_{\rm B} \sim \rho c_g v_{\rm K}.
\end{equation}
In this limit the magnetic pressure exceeds the gas pressure by a factor 
\begin{equation}
\label{beta}
\beta^{-1} \equiv {p_{\rm B} \over p_g} \approx {v_{\rm K} \over c_g} \gg 1;
\end{equation}
thus, magnetic pressure would dominate the structure of the disc.

\subsection{MRI and the Limiting Toroidal Field Strength}

In this section, we analyze the physics behind the limiting field
strength, and show its relationship to other stability results in the
literature.  Pessah \& Psaltis (2005) consider a differentially
rotating, cylindrical equilibrium flow with no $z$-dependence. In
their analysis, they make approximations that correspond to assuming
that the flow is of uniform density and contains a uniform poloidal
field $B_z \neq 0$ and a uniform toroidal field $B_\phi$ (their
results can be generalized to include a radial gradient of $B_\phi$).
The angular velocity is taken to be of the form $\Omega \propto
R^{-q}$. They consider only axisymmetric perturbations, and this
enables them to carry out a local stability analysis and to obtain a
local dispersion relation relating wave frequency $\omega$ and
wavevector ${\bf k} = (k_R, 0 , k_z)$, in cylindrical polar
coordinates $(R, \phi, z)$. Their analysis differs from previous work
in that they take account of both compressibility and geometrical
terms. They focus on the case $k_z \gg k_R$ in order to study how the
MRI modes are affected by the strength of the azimuthal field.

In such a configuration, one might expect to encounter two different
types of instability: (i) the MRI, which uses the magnetic field (here
the poloidal component, which is why we require $B_z \neq 0$ and $k_z
\gg k_R$) to tap the energy in the shear, and (ii) buoyancy-driven
instability, which makes use of the radial structure of the
equilibrium configuration to tap the (radial) effective gravity.
Pessah \& Psaltis find three regions of instability. Regions I and II
(see their Figures 3 and 5) seem to correspond to the two types of
instability mentioned above.

Type I instability occurs at small $k_z$ and cuts off at a finite
value of $k_z = k_{BH}$. This corresponds roughly to modes being
unstable only when the time for a vertical magnetic wave, with
wavespeed $v_{Az} = B_z/\sqrt{4\pi \rho}$, to cross one (vertical)
wavelength ($\lambda \sim 1/k_z$) is longer than the orbital time
$1/\Omega$. This is the standard criterion for instability to MRI, and
corresponds physically to the situation in which a vertical field line
is unable to straighten itself fast enough to counter rotational
effects (centrifugal force).  Thus instability requires
\begin{equation}
\label{cond1}
k_z v_{Az} \la \Omega.
\end{equation}

Pessah \& Psaltis also find that the MRI modes are stabilized if the
azimuthal field is strong enough. They find that modes of Type I are
unstable only for low values of $B_\phi < B_{\rm crit}$ such that
\begin{equation}
\label{PPcriterion}
v_{A\phi}^2 \la v_K c_s.
\end{equation}
Here, as before, $v_{A \phi} = B_\phi /\sqrt{4\pi\rho}$ is the
azimuthal Alfv\'en speed, $v_K$ is the Keplerian azimuthal velocity
and $c_s$ is the sound speed. The regime we are interested in
corresponds to a suprathermal, strongly azimuthal field, so that $ c_s
\ll v_{{\rm A}\phi} \ll v_K$ and $B_z \ll B_\phi$. This regime differs
from the situation in which the predominant field is $B_z$ in two
important respects. First, the azimuthal field exerts an additional
restoring force for a $k_z$ perturbation, which, for a given $k_z$,
dominates at large $B_\phi$. Second, the length of a field line
corresponding to one vertical wavelength ($\lambda \sim 1/k_z$) is now
\begin{equation}
d \sim \lambda \frac{B_\phi}{B_z}.
\end{equation}

The analysis of Pessah \& Psaltis appears to suggest that the MRI is
no longer able to operate when the azimuthal field is increased to the
extent that the timescale on which the restoring force due to the
field operates ($\sim R/v_{A\phi}$) is shorter than the timescale on
which pressure equilibrium can be established along the field line
($\sim d/c_s$). In this case, instability requires
\begin{equation}
\label{cond2}
k_z v_{Az} \ga \frac{v_{A\phi}^2}{c_s R}.
\end{equation}
Putting conditions (\ref{cond1}) and (\ref{cond2}) together, we obtain
the condition for instability given by inequality (\ref{PPcriterion})
above.

Pessah \& Psaltis suggest that Type II instability corresponds to
buoyancy modes, in line with the results of Kim \& Ostriker (2000) for
the limit $\cs \rightarrow 0$. Modes of Type I and Type II are clearly
linked in some way, as they undergo an exchange of stability at the
critical azimuthal field strength. Type II modes are unstable only
when both inequalities (\ref{cond1}) and (\ref{cond2}) are violated,
and this only occurs for large values of the azimuthal field strength
$B_\phi > B_{\rm crit}$.  Thus, for a given vertical field strength
$B_z$, MRI modes --- which tap the shear energy --- are able to
operate for small $k_z$ and $B_\phi < B_{\rm crit}$, whereas for
$B_\phi > B_{\rm crit}$, radial buoyancy modes, with large $k_z$, take
over.
 
The derivation of the stability criterion by Pessah \& Psaltis
requires the presence of a vertical field ($B_z \neq 0$), however
small. But analogous behavior is found for the case in which the field
is purely azimuthal (Terquem \& Papaloizou 1996). Since MRI draws its
energy from the background shear, it is evident that instability can
arise in this case only for nonaxisymmetric modes. Terquem \&
Papaloizou use trial local displacements to show that the linear
operator describing the evolution of the linearized equations has a
dense or continuous spectrum of oscillation frequencies $\omega$ that
satisfy a local dispersion relation.  Their analysis is restricted to
modes that are essentially incompressible in the
$(R,z)$-plane.\footnote{Note that this restriction still permits the
effects of compressibility to operate in the azimuthal direction.}
This eliminates the fast MHD mode, which in any case is not involved
in MRI, and means that the dispersion relation is only a quartic in
$\omega$.

In the limit we are considering, with $k_z \gg k_R$ and $c_s \ll
v_{\rm A} \ll v_K$, the Terquem \& Papaloizou dispersion relation
simplifies considerably. For axisymmetric ($m=0$) disturbances,
$\omega$ is real, implying stability.  For non-axisymmetric
disturbances, we consider first modes with $0 < m \la v_K/ v_{\rm A}$.
For these modes the timescale for an azimuthal magnetic wave, with
wavespeed $v_{\rm A}$, to cross one (azimuthal) wavelength ($\sim
R/m$), is longer than the orbital time $1/\Omega$ (cf. condition
[\ref{cond1}]).  For this range of $m$ there are two low frequency
roots which can be unstable (in addition to two stable roots with
$\omega + m \Omega \approx \pm \Omega$). What is interesting for our
current discussion is that the criterion for instability corresponds
approximately to condition (\ref{PPcriterion}) derived by Pessah \&
Psaltis. The growth rates are of order $\sim m c_s/R$, and the (real
parts of the) frequencies in the rotating frame are such that $\omega
+ m \Omega
\sim m v_{\rm A}^2 / R^2 \Omega \la v_{\rm A}/R$.

For higher values of $m$, such that $m \ga v_K/v_{\rm A}$, one finds
instability provided that $m \la v_A/c_s$. Putting these two
conditions together thus implies that for instability we require
$v_{\rm A}^2 > c_s v_K$, which is the reverse of condition
(\ref{PPcriterion}). This may correspond to the exchange of
stabilities found by Pessah \& Psaltis between their modes of Type I
and Type II. The growth rates for these high $m$ modes are $\sim
v_{\rm A}/R$, and the real parts of the frequencies in the rotating
frame are small, $\omega + m \Omega \sim \Omega/m$. The fact that
these modes are almost stationary in the corotating frame and that the
growth rates depend on the field strength suggests, as for the Pessah
\& Psaltis Type II modes, that they are driven by buoyancy.
 
The analogy between the findings of Pessah \& Psaltis for axisymmetric
modes in the limit of $0 < B_z \ll B_\phi$, and the analysis of
Terquem \& Papaloizou for nonaxisymmetric modes when $B_z = 0$, seems
to imply that the same physical processes are at work in both cases.

The basic conjecture of this paper is that amplification of the
toroidal magnetic field occurs only while MRI is available to drive a
dynamo.  Since MRI dies out as $B_{\rm crit}$ is approached from
below, we assume that $B_\phi$ saturates at roughly this
level. Indeed, the growth rates of both Type I and Type II modes are
small in the vicinity of $B_{\rm crit}$, where the exchange of
stability between MRI and buoyancy modes occurs.  However, even if
$B_\phi$ managed to cross this bottleneck and buoyancy modes were
excited, this does not mean that $B_\phi$ will continue to grow.
Since buoyancy modes do not tap the shear, they are unlikely to drive
a dynamo capable of amplifying the toroidal magnetic field.

\section{Consequences for Disc Structure}

In this section we develop a one-zone model for vertical disc
structure, under the assumption that the toroidal field reaches its
limiting strength according to the Pessah \& Psaltis (2005) analysis.
The limiting field strength is related to the ability of sound waves
to propagate along toroidal field lines.  Because of the large
radiative diffusivity of accretion discs, such waves are unlikely to
be mediated by radiation pressure, even where radiation dominates the
total pressure.  Therefore we assume that the appropriate sound speed
to use in estimating $\va$ is that due to the gas pressure, $\cg \sim
(p_g / \rho)^{1/2}$, and we take $\va \sim (\cg \vk)^{1/2}$. For a
thin disc we have the ordering $c_g \ll \va \ll \vk$.  When $\va$
exceeds the sound speed associated with the radiation pressure, $c_r
\sim (p_r/\rho)^{1/2}$, the scale height is given by
\begin{equation}
\label{hr}
{H\over R} \sim {\va \over \vk} \sim \left({c_g\over \vk}\right)^{1/2},
\end{equation}
instead of the Shakura--Sunyaev result, $H_{\rm SS}/R \sim \cs /\vk
=(c_g^2 + c_r^2)^{1/2}/\vk$ (eq.~\ref{honr}).  If we assume an
$\alpha-$model viscosity with kinematic viscosity given by $\nu =
\alpha H \va$ (rather than the usual $\nu = \alpha H c_s$, but
preserving the assumption that the stress is $\sim \alpha p_B$ with
$p_B$ being essentially the total pressure), then the inflow speed
(for a Keplerian disc) is given by
\begin{equation}
\label{vin}
v_{\rm in} \sim {3\over 2}\alpha \left({H\over R}\right)^2 \vk \sim
{3\over 2} \alpha \cg
\end{equation}
and the column density is
\begin{equation}
\label{Sigma}
\Sigma \sim {\dot M \over 3\pi \alpha \cg R} ,
\end{equation}
where $\dot M$ is the local mass accretion rate through the disc.

It is convenient to normalize radii to the gravitational radius, $R_g
= GM/c^2$, and the accretion rate to the Eddington accretion rate,
which we define as $\me = \le/c^2 = 4\pi GM/ \kappa c$, where $\kappa$
is the opacity and $M$ is the mass of the central object.  Defining
$\dot m \equiv \dot M / \me$ and $x \equiv R/R_g$, we obtain an
expression for the transverse optical depth through the disc,
\begin{equation}
\label{tau}
\tau = \Sigma\kappa \sim {4\over 3}{\dot m \over \alpha}{c\over \cg} x^{-1}.
\end{equation} 
If the disc is radiative, the flux from each side (at $x \gg x_{\rm
in}$, where $x_{\rm in} = 6$ is the radius of the innermost stable
orbit for a Schwarzschild black hole) is given by
\begin{equation}
\label{F}
F \approx {3\over 8\pi}{GM \dot M \over R^3} = {3\over 2} {c^5\over
GM\kappa }{\dot m \over x^3} \sim {2 p_r c \over \tau} ,
\end{equation}
where $p_r$ is the radiation pressure near the midplane and the factor
2 in the last relation arises because the flux escaping from each side
of the disc traverses half the optical depth (which is assumed to be
$> 1$).  We now solve for the radiation pressure,
\begin{equation}
\label{pr}
p_r \sim {c^4\over GM\kappa }{c\over \cg} {\dot m^2 \over \alpha} x^{-4}.
\end{equation}
The density inside the disc is given by
\begin{equation}
\label{rho}
\rho = {\Sigma \over 2 H} \sim {2\over 3}{c^2\over GM\kappa} \left({c\over \cg}\right)^{3/2} {\dot m \over \alpha} x^{-9/4}.
\end{equation}

The results presented so far are fully general.  To obtain numerical
results from them, however, one must determine the value of $\cg$ and
check whether $\va > c_r$. This depends on the equation of state, the
opacity, and the relative importance of radiation and gas pressure,
which we now address.

\subsection{The inner regions of black hole accretion discs}

\subsubsection{The disc structure}

We consider here the inner regions of a standard Shakura--Sunyaev disc
around a black hole where radiation pressure dominates over gas
pressure at small radii for values of $\dot m$ that are not too small.
The opacity is dominated by electron scattering and we take $\kappa =
\kappa_{es} = 0.4$ cm$^2$ g$^{-1}$.  We begin by calculating the
magnetically dominated disc structure under the assumption $\pb >
p_r$.  We then check the self-consistency of this assumption {\it a
posteriori}, to find the radius within which $p_r > \pb$ and magnetic
support can be neglected.

To determine both $\cg$ and $p_r$, we assume that the radiation inside
the disc is in local thermal equilibrium (LTE) and set $T= T_{\rm LTE}
= (3 p_r/a)^{1/4}$.  We then use eq.~(\ref{pr}) to solve for $T$ with
$c_g = (kT/\mu)^{1/2}$, where $\mu \approx 0.6 m_p$ is the mean mass
per particle.  We verify the LTE assumption {\it a posteriori}.  We
find that $\va > c_r$ for $x > 20 (\alpha m)^{2/37} \dot m^{32/37}$,
implying that magnetic pressure dominates over other forms of pressure
at all radii, except possibly for the region immediately outside a
black hole or neutron star.  Radiation pressure dominates over gas
pressure for $x < 2.0 \times 10^3 (\alpha m)^{2/13} \dot m^{8/13}$,
but since magnetic pressure determines the vertical structure through
most of this region, there is no difference in properties between the
radiation and gas pressure-dominated zones.

The electron scattering optical depth to the disc midplane is 
\begin{equation}
\label{taues}
\tau_{\rm es} \sim 2.5 \times 10^2 \alpha^{-8/9} m^{1/9}\dot m ^{7/9} x^{-5/9} .
\end{equation}
To check the assumption of LTE, we also need the absorption opacity,
which we take to be the Kramers bound-free opacity for solar
abundances, $\kappa_{\rm bf} = 1.6\times 10^{24} \rho T^{-7/2}$ cm$^2$
g$^{-1}$.  The effective optical depth for thermalization is then
\begin{equation}
\label{taustar}
\tau_* = \tau_{\rm es} \left({\kappa_{\rm bf} \over \kappa_{\rm es} }\right)^{1/2} \sim 0.29 \alpha^{-11/12} m^{1/12} \dot m^{1/3} x^{5/24} ,
\end{equation}
indicating for typical values of $\alpha$ that in general, LTE is a
reasonable approximation.
 
Finally, we estimate the disc thickness.  Substituting into
eq.~(\ref{hr}), we obtain
\begin{equation}
\label{hrg}
{H \over R} \sim 7.3 \times 10^{-2}(\alpha m)^{-1/18} \dot m^{1/9}
x^{1/36} \ {\rm cm^2} \ {\rm g}^{-1} .
\end{equation}
The disc aspect ratio, $H/R$ (the ``opening angle"), is nearly
independent of radius and is extremely insensitive to parameters, with
a magnitude several times larger than a comparable Shakura--Sunyaev
disc without magnetic support.

\subsubsection{Thermalization and Colour Temperature}

Photospheric colour temperatures typically exceed effective
temperatures in the inner regions of standard accretion discs because
the radiation is thermalized at a significant scattering optical
depth.  Colour corrections of luminous accretion discs, $f_{\rm col}
\equiv T_{\rm col}/ T_{\rm eff}$, usually lie in the range $\sim
1.5-2$ (Shimura \& Takahara 1995; Davis et al. 2005), corresponding to
a scattering optical depth $\sim 5-20$ at the thermalization layer.
When $\tau_* > 1$, the colour correction is given roughly by $f_{\rm
col} \sim (\kappa_{\rm es} / \kappa_*)^{1/8}$, where $\kappa_*$ is the
absorption opacity evaluated at the local disc colour temperature and
at the density of the thermalization layer, $\rho_*$.

An accurate calculation of the colour correction requires a detailed
model for the structure of the disc photosphere, which is beyond the
scope of our one-zone model for the vertical structure.  However, it is
clear that a magnetically supported disc should have a higher colour
temperature than the equivalent Shakura--Sunyaev model (Blaes et
al.~2006).  Here we make a crude estimate of this effect.

The absorption opacity at the thermalization layer is given by 
\begin{equation}
\label{kappastar}
\kappa_* = \kappa_{\rm bf}(\rho_*, T_{\rm col}) = \kappa_{\rm bf} {\rho_* \over \rho} \left({T\over T_{\rm col}} \right)^{7/2}, 
\end{equation}
where the unsubscripted density and temperature represent values at
the midplane.  We also have $T/T_{\rm col} = (T_{\rm eff}/ T_{\rm
colour}) (T/T_{\rm eff} ) = \tau_{\rm es}^{1/4} f_{\rm col}^{-1}$.
Substituting for the temperature ratio in eq.~(\ref{kappastar}) and
using eq.~(\ref{taustar}), we can solve for $f_{\rm col}$ to obtain
\begin{equation}
\label{fc}
f_{\rm col} \sim { \tau_{\rm es}^{1/4}\over \tau_*^{4/ 9} }
\left({\rho_* \over \rho}\right)^{-2 / 9} \ga 3.4 \alpha^{5/27}
m^{-1/108}\dot m^{5/108} \left({x\over 20} \right)^{-25 / 108} ,
\end{equation}
since $\rho_* / \rho < 1$. In the inner portions of the magnetically
dominated zone, eq.~(\ref{fc}) predicts a colour correction
substantially larger than that of a standard disc.

\subsubsection{Thermal and Viscous Stability}

A major problem with standard $\alpha$-disc models is that
they are viscously and thermally unstable in their inner regions where
electron scattering dominates the opacity (Lightman \& Eardley 1974;
Shakura \& Sunyaev 1976; Pringle 1976). In contrast, the magnetically
supported disc models we propose here are not subject to these thermal
and viscous instabilities. In all radiative $\alpha-$disc models in
which the opacity is dominated by electron scattering, the dissipative
heating rate scales as $Q^+ \propto H^2 \Sigma$ and the radiative loss
rate scales as $Q^- \propto p_r/\Sigma \propto T^4 / \Sigma$, where we
assume LTE. From eq.~(\ref{hr}), we find $T \propto H^4$ at fixed $R$,
implying that $Q^- \propto H^{16}/ \Sigma$.  At constant $\Sigma$,
losses increase with $H$ much more rapidly than heating, implying that
the magnetically supported discs are thermally stable.  This result
can be contrasted to a radiation pressure-dominated $\alpha$-disc, in
which $T\propto H^{1/4}$, implying instability. (Gas
pressure-supported $\alpha$-discs, on the other hand, are thermally
stable.)

Now consider viscous instability. On viscous timescales, thermal
balance is maintained, implying $Q^+ = Q^-$.  Therefore, $\Sigma
\propto H^7$ for the magnetically supported disc.  The viscous couple
satisfies $G \propto H^2 \Sigma \propto \Sigma^{9/7}$.  Since the
viscous couple is an increasing function of surface density, the disc
is stable (Lightman \& Eardley 1974; Pringle 1981).
 
\subsection{The thickness of CV and LMXB discs}

By using the same analysis as above, but now using Kramers, rather
than electron scattering, opacity, we can calculate the thickness of
the outer regions of discs in binary stars. As a specific example, we
consider a disc around a 1 M$_\odot$ white dwarf, at a
radius of $R = 10^{10} R_{10}$ cm, accreting at a rate of $\dot{M} =
10^{18} \dot{M}_{18}$ g s$^{-1}$ $\approx 10^{-8}$ M$_\odot$
yr$^{-1}$. This corresponds to the outer regions of the disc of a dwarf nova 
in outburst or of a nova-like variable. Under the standard assumptions
the thickness of such a disc would be $H/R \sim 0.05$ (see,
for example, Frank, King \& Raine 2002). However, if we use the
assumption of strong toroidal fields put forward in this paper we find
that the disc thickness is given by
\begin{equation}
\label{CVdiscH}
\frac{H}{R} = 0.48 \alpha^{-1/17} \dot{M}_{18}^{3/34} \left(
\frac{M}{\msun} \right)^{-15/68} R_{10}^{9/68}.
\end{equation}

From this we see that not only are the strongly magnetized discs
proposed here thicker than standard ones by about an order of
magnitude, but they are also slightly more flared, with $H/R \propto
R^{0.132}$ rather than the usual $\propto R^{0.125}$. Thus such discs are more
subject to irradiation from a central luminosity source.

\subsection{Reduction of Disc Self-Gravity}

Magnetic pressure support can reduce the effects of self-gravity in
the outer parts of an accretion disc by thickening the disc and
reducing its density (Pariev et al.~2003).  For a disc supported by
gas pressure, local gravitational instability and fragmentation are
expected to limit the accretion rate to
\begin{equation}
\label{shlos} 
\dot M < \dot M_{\rm max} \sim {3\alpha c_g^3 \over G} = 5\times 10^{-4} \alpha T_{100}^{3/2} M_\odot \ {\rm yr}^{-1},
\end{equation}
where $T_{100} = T/100$ K (Shlosman \& Begelman 1987).  In a galactic
nucleus, far from the black hole, the disc temperature is set by
environmental influences (external irradiation, cosmic rays, etc.),
rather than the internal dissipation in the disc (Shlosman \& Begelman
1989).

The maximum accretion rate given by eq.~(\ref{shlos}) is too small to
power luminous active galactic nuclei (AGN). Magnetic pressure
support, according to our prescription, would increase this upper
limit by a factor $(\vk / \cg)^{3/2} \approx 10^5 (M_9/ T_{100}R_{\rm
pc})^{3/4}$, where black hole mass and radius are in units of $10^9
M_\odot$ and pc, respectively. In Eddington units, the maximum
accretion rate that could pass through a radius $R$ would be increased
to
\begin{equation}
\label{shlos2} 
\dot m_{\rm max} \sim 25 \alpha M_9^{-1/4} T_{100}^{3/4} R_{\rm pc}^{-3/4}   .
\end{equation}
The decline of $\dot m_{\rm max}$ with $R$ continues out to a radius
$R_{\rm BH}$, where the gravitational potential of the galaxy is
comparable to that of the black hole.  If the galactic nucleus is
represented by a singular isothermal sphere with velocity dispersion
$\sigma$, the ``bottleneck" occurs at $R_{\rm BH} \sim GM/2\sigma^2$
--- beyond $R_{\rm BH}$, the carrying capacity of the disc, at fixed
$T$, is independent of radius.  If we use the $M-\sigma$ relation
(Ferrarese \& Merritt 2000; Gebhardt at al. 2000; Tremaine et
al. 2002), $M_9 \sim 0.13 (\sigma/ 200 \ {\rm km \ s}^{-1})^4$, to
eliminate $\sigma$ in favor of $M$, we obtain
\begin{equation}
\label{shlos3} 
\dot m_{\rm max} \sim 3 \alpha M_9^{-5/8} T_{100}^{3/4}   .
\end{equation}
Thus, magnetic support may permit accretion at close to the Eddington
limit, provided that $\alpha$ and $T_{100}$ are not too small. (Note
that we defined $\dot M_{\rm Edd}$ without an efficiency factor, so
$\dot m \sim 10$ is required to produced an Eddington luminosity with
an efficiency of 0.1.)

\section{Discussion and Conclusions} 

We are well-aware of the speculative nature of some of the assumptions
that have gone into our proposed disc model.  Our proposal that
$v_{\rm A}$ saturates at $\sim (\cs \vk)^{1/2}$ is an educated guess
based on a very simple interpretation of the likely conditions under
which MRI is able to drive strong turbulence. We implicitly assume
that MRI operates at all heights above the disc midplane, amplifying
$B_\phi$ in situ; however, $B_\phi$ may also be advected to high
latitudes by buoyancy.  If buoyant transport from below dominates,
then $v_{\rm A}$ could be even larger than we suggest.

The surviving MRI modes when $v_{\rm A} \gg c_s$
have large $k_z$, implying that the resulting turbulence would be
driven on small scales.  This is in contrast to the more common
assumption in MRI calculations, that $v_{\rm A} \ll c_s$, in which
case turbulence is driven on scales up to the disc thickness.  In
order to build up the strong, large-scale toroidal field in our
picture, the dynamo might have to involve an inverse cascade, the
possibility of which is by no means certain.  If the production of a
large scale toroidal field is not efficient in this limit, then our
proposed field strength could be a considerable overestimate.

Alternatively, we have suggested the possibility that accretion discs
with strong toroidal fields of the kind we discuss here may form a
second stable branch of accretion disc configurations, perhaps only accessible through some sudden change in properties --- for example, a thermal instability (cf.~Section 3).  This kind of bimodality of thin discs could help to explain the spectral differences of otherwise similar accreting white dwarf systems in CVs (cf.~Section 2.1).

Whether or not our quantitative estimate of the saturated field
strength is accurate and physically realized, we wish to emphasize the
role that a magnetic ``equation of state" --- a relation between a
suprathermal $B_\phi$ and other fluid variables --- could play in
altering the structure and stability properties of $\alpha-$discs.
 
\section{Acknowledgments} 

We thank Keith Horne, Frank Verbunt, Richard Wade and Gordon Ogilvie
for helpful discussions. MCB acknowledges support from NSF grant
AST-0307502 and from the University of Colorado Council on Research
and Creative Work. He thanks the Institute of Astronomy and Trinity
College, Cambridge for hospitality. JEP thanks the Fellows of JILA for
their hospitality.

\end{document}